# An HIV Feedback Resistor: Auto-Regulatory Circuit Deactivator and Noise Buffer

**Leor S. Weinberger**\*, **Thomas Shenk**

Department of Molecular Biology, Princeton University, Princeton, New Jersey, United States of America

Animal viruses (e.g., lentiviruses and herpesviruses) use transcriptional positive feedback (i.e., transactivation) to regulate their gene expression. But positive-feedback circuits are inherently unstable when turned off, which presents a particular dilemma for latent viruses that lack transcriptional repressor motifs. Here we show that a dissipative feedback resistor, composed of enzymatic interconversion of the transactivator, converts transactivation circuits into excitable systems that generate transient pulses of expression, which decay to zero. We use HIV-1 as a model system and analyze single-cell expression kinetics to explore whether the HIV-1 transactivator of transcription (Tat) uses a resistor to shut off transactivation. The Tat feedback circuit was found to lack bi-stability and Tat self-cooperativity but exhibited a pulse of activity upon transactivation, all in agreement with the feedback resistor model. Guided by a mathematical model, biochemical and genetic perturbation of the suspected Tat feedback resistor altered the circuit's stability and reduced susceptibility to molecular noise, in agreement with model predictions. We propose that the feedback resistor is a necessary, but possibly not sufficient, condition for turning off noisy transactivation circuits lacking a repressor motif (e.g., HIV-1 Tat). Feedback resistors may be a paradigm for examining other auto-regulatory circuits and may inform upon how viral latency is established, maintained, and broken.

## Introduction

Many viruses encode strong transcriptional positive-feedback (i.e., excitatory-feedback) loops in order to quickly ramp up their gene expression after invading the host cell. But positive-feedback loops are inherently unstable when turned off (henceforth referred to as the "off state"), which frustrates a virus' ability to maintain a stable latent/lysogenic state. Thus, viruses that maintain latent or lysogenic states (e.g., lentiviruses, herpesviruses, and bacteriophages) face an inherent design dilemma.

A transcriptional off state is a necessary, but possibly not sufficient, condition for establishing viral latency. Therefore, latency-prone viruses that have retained the advantages of positive feedback must have evolved control methods to stabilize their transcriptional off states. Below we present a model where covalent modification and back-modification of the positive-feedback "transactivator" introduces nonfunctional, reversible intermediate states into the positive-feedback loop. Such covalent interconversion (also called a "futile cycle" [1]) is analogous to introducing a dissipative (i.e., nonadiabatic) resistor into an electrical feedback circuit. The dissipative resistor can stabilize the off state of a positive-feedback loop.

The bacterial virus, phage-λ, is the archetypal model to explain how viruses overcome the feedback-versus-stability dilemma and maintain latency (reviewed in [2]). λ can actively replicate (its lytic state) or it can enter quiescence (lysogeny). The choice between growth states is made by a circuit composed of a bi-directional promoter region that controls the expression of λ repressor, which favors lysogeny, and Cro, which initiates the lytic cycle of infection. Cooperative binding of λ repressor to operator motifs within the bi-directional promoter favors the synthesis of mRNA that encodes repressor and inhibits the synthesis of mRNA that encodes Cro (Figure 1A). Cooperativity, which results from oligomerization of λ repressor, is critical to the function of the circuit. When the repressor is abundant, it forms octamers that directly bind to operator sequences and inhibit further phage-λ gene expression. However, when the repressor concentration is lowered (e.g., following activation of the host SOS system in response to DNA damage) it can no longer oligomerize nor bind to operators. This reduction allows Cro to be activated, initiating lytic replication. λ repressor oligomerization and cooperativity thereby produce a circuit that is bi-stable and can maintain either a lytic or a lysogenic state [3].

Animal virus regulatory circuits operate differently than λ does. Many animal viruses can enter latency, but are not known to have bi-stable λ-like repressor motifs (Figure 1B). Instead, latent animal viruses appear to be regulated by relatively simple transactivation motifs (i.e., excitatory feedback). It is a mystery how viruses lacking bi-stable λ-like repressor motifs and encoding strong positive feedback can establish stable off states, which are at least necessary for latency. Cooperativity in these positive-feedback loops (dependent on oligomerization of the transactivator) can in

## Author Summary

Many viruses have the cunning ability to enter a hibernative or off state, termed latency. When in a latent state, the virus is unable to replicate, and its gene expression program is largely shut down. This facility for lying dormant typically ensures lifelong persistence of the virus in the host; it is also a particularly problematic obstacle in the treatment of HIV. For most viruses, the molecular regulation of entry into latency is not completely understood, but it is believed that viral gene expression must be deactivated in some way. In this study, we introduce a new regulatory motif, the feedback resistor, that enables a genetic circuit to shut off without the need for an active repressor molecule. We first show that many animal viruses might encode feedback resistors in their regulatory circuits. Then, by using a combination of mathematical theory and single-cell real-time imaging experiments, we show that a feedback resistor in the HIV Tat transcriptional circuit likely allows the HIV virus to enter into latency. We postulate that feedback resistors may give increased stability and control in the complex noisy signaling environment of the cell.

principle maintain a stable off state [4,5], but below, we examine a circuit that appears to lack cooperativity.

Herpesviruses, in particular, establish and maintain latency without utilizing a bi-stable λ-like repressor motif. For example, Epstein Barr virus (EBV) encodes the Rta transactivator. Rta establishes an excitatory-feedback loop [6], and it induces the expression of other virus genes through multiple mechanisms, including the viral transactivator protein Zta [7]. Expression of Rta has been shown to disrupt EBV latency [8], so a relatively stable Rta transactivator off state is presumably necessary for maintaining latency. Similarly, in herpes simplex virus type 1 (HSV-1), the ICP0 protein plays the central role in the induction of lytic replication and reactivation from latency [9]. ICP0 transactivates its own promoter and activates viral gene expression through multiple mechanisms [10]. Although there is evidence that ICP0 might be inhibited post-transcriptionally by the LAT RNAs (these anti-sense RNAs have the potential to bind ICP0 mRNA and are expressed in latently infected cells [11]), post-transcriptional LAT inhibition of ICP0 is only partial, because latently infected cells lack detectable ICP0 mRNA. Thus, there is likely a transcriptional component to ICP0 silencing as well. Furthermore, there is currently no evidence for bi-stable λ-like repressor motifs in EBV or HSV-1. Likewise, cytomegalovirus (CMV) also apparently lacks a bi-stable λ-like repressor motif. CMV lytic replication is controlled by the major immediate-early circuit, which encodes the IE1 and IE2 proteins via alternative splicing. Both of these proteins are vital transactivators [12,13]; IE1 transactivates the major immediate-early promoter and IE2 is essential for activating viral early and late genes [14,15]. IE2 also partially down-modulates IE1/IE2 expression [16,17], but because IE2-mediated down-modulation is partial, there is likely a additional component to major immediate early silencing as well.

Thus, underlying all the regulatory complexity, each animal virus regulatory circuit appears to be built on the scaffold of a positive-feedback loop, and despite that these regulatory circuits are modulated by numerous viral and cellular factors, there is no clear understanding of how a transcriptional off state is achieved. Although host factors undoubtedly contribute importantly to the function of viral transactivator circuits, virus-coded elements appear to run the basic excitatory-feedback programs that control the majority of transactivation kinetics. Specifically, in the case of HIV-1 transactivator of transcription (Tat), CDK9 is constitutively expressed throughout the cell cycle [18], and the levels of CDK9/CyclinT1 heterodimer (an essential Tat co-activator) remain constant [19]. Furthermore, host mechanisms such as chromatin modification are thought to be involved mainly with maintenance of repression [20,21], rather than active repression of ongoing high-level transcription. Specifically, the removal of active transcriptional machinery (i.e., an off state) appears to be a prerequisite for the formation of repressive chromatin, as highlighted by studies on the HSV thymidine-kinase promoter [22]. Thus, it is necessary to understand how viral transactivation circuits establish off states and what the key regulators might be.

HIV-1 is an example of a latent virus lacking a λ-like repressor and encoding an excitatory transactivator (Figure 1B. HIV-1 establishes a quiescent or latent state in CD4+ T cells, and the choice between replication versus latency is controlled by HIV-1-encoded Tat protein. (reviewed in [23]). Tat enhances transcriptional elongation from the HIV-1 promoter, the long terminal repeat (LTR), by facilitating CDK9 mediated hyperphosphorylation of RNA polymerase II (RNAPII) and increasing elongation processivity. It is now clear that Tat transactivation involves two distinct steps. First, deacetylated Tat binds to the nascent TAR RNA loop in the LTR, recruits the positive transcriptional elongation factor b (pTEFb), and is then acetylated by the host histone acetylatranferase p300. Second, acetylated Tat recruits the host SWI/SNF chromatin-remodeling complex to the LTR, thereby facilitating remodeling of a single nucleosome (nuc-1) and allowing the highly processive RNAPII to proceed downstream [24]. Tat acetylation appears to be a rate-limiting step, because deacetylated Tat is in great excess in the cell [25]. Tat ultimately up-regulates its own gene expression 50–100-fold above basal levels and concurrently up-regulates the expression of its alternative splice variant Rev. Rev forms a multimer that is essential for replication and exports unspliced HIV mRNA (e.g., genomic mRNA) from the nucleus [26]. Rev concentrations must reach a critical threshold for efficient Rev multimer formation, and Tat must drive Rev concentrations above this threshold for lytic replication to ensue. Despite the many host factors that are implicated (and likely required) in HIV-1 proviral latency, at the very least, a Tat off state appears necessary for establishing HIV-1 proviral latency [27,28] . It is unclear how the Tat positive-feedback loop can establish such an off state.

In principle, either a λ-like repressor or cooperative positive feedback could induce bi-stability in the HIV-1 Tat circuit and stabilize the off state. But, no HIV-1 encoded self-repressors are known, and evidence for Tat cooperativity is mixed. Whereas early studies showed that purified Tat protein could form metal-linked dimers in a highly reducing in vitro environment [29], later studies found that Tat dimers fail to activate transcription [30]. Also, Tat appears to be a monomer in cells [31,32], and Tat monomers are required for transactivation [33]. Recent biochemical analyses are consistent with Tat having a stoichiometry equal to 1 in the transcriptional elongation complex in vivo [25]. Without oligomerization or cooperativity, the Tat circuit presents us

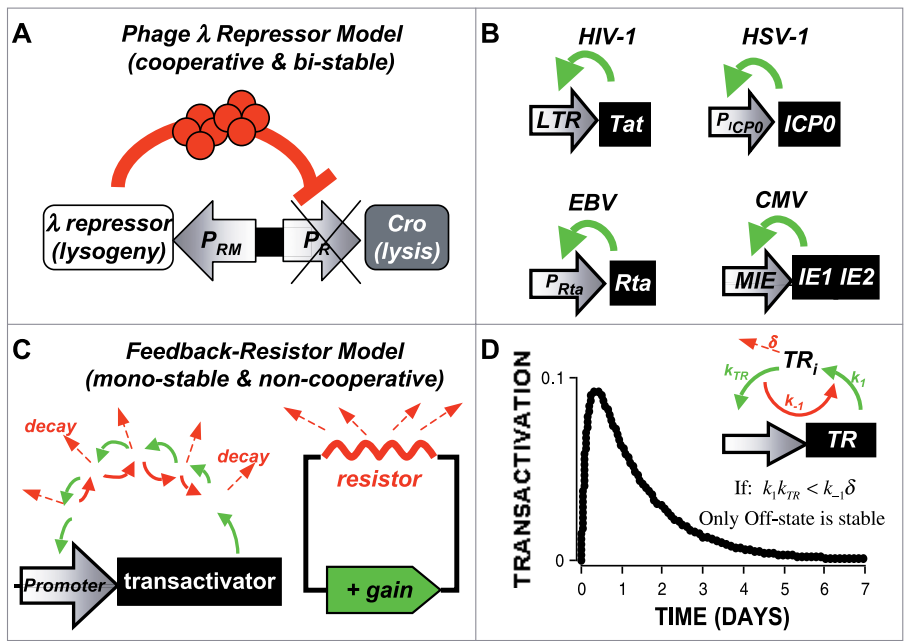

**Figure 1.** Repressor Versus Resistor: Latent Animal Virus Circuits Lack λ-Like Bi-Stable Repressor Motifs but Might Maintain Stability Using a Feedback Resistor

(A) Cartoon of the cooperative, bi-stable λ repressor model. λ DNA contains a bi-directional operator/promoter and λ repressor forms an octamer that binds to operator elements and inhibits Cro expression. Blocks represent genes and faded block arrows represent respective promoters $P_R$ and $P_{RM}$.

(B) Four animal viruses lack bi-stable repressor motifs in their regulatory circuits and encode simple transactivation motifs. Blocks represent genes and faded block arrows represent promoters.

(C) The feedback resistor model. Enzymatic interconversions (a futile cycle) of the transactivator to its final functional form generate nontransactivating intermediates and make up a dissipative (or nonadiabatic) resistor in an excitatory feedback circuit. Green arrows are forward modifications, red arrows are a reversal of the modification, and dashed red arrows are decay (turnover).

(D) The feedback resistor generates a pulse of transactivation. For simplicity, we present a numerical simulation of the feedback resistor model for a transactivator $(Tr_I)$ having a single intermediate $(Tr_1)$: $dTr_I/dt = -k_1Tr + (k_{-1} + k_{TR})Tr_I - \delta Tr$; $dTr_1/dt = k_1Tr - k_{-1}Tr_I$ when $k_1k_{TR} < k_{-1}\delta$ (i.e., all eigenvalues are negative) and initial condition $Tr(t=0) > 0$. If the product of the backward reaction rates (i.e., negative reaction rates $k_{-1}$ and $\delta$) exceeds the forward modification rates (i.e., positive reaction rates $k_1$ and $k_{TR}$), then a small amount of transactivator at time $t = 0$ generates a burst of transactivation that eventually decays over time to the only stable state: the off state $(Tr,Tr_1) = (0,0)$. Under these conditions, this stable off state is attracting (mathematically, all eigenvalues are real and negative, because the Jacobian for this system has negative trace and positive determinant). In general, increasing the number of dissipative intermediate states $(Tr_I)$ reduces the duration of the pulse, whereas increasing the number of nondissipative intermediate states generates a gamma-distributed delay and lengthens the duration of the pulse.

doi:10.1371/journal.pbio.0050009.g001

with a positive feedback–versus-latency dilemma. Moreover, we have recently shown that the Tat circuit is sensitive to stochastic molecular fluctuations intrinsic to gene expression but also maintains an off state despite these fluctuations [34]. In a stochastic environment, random molecular noise should routinely perturb an excitatory-feedback loop such as Tat away from the unstable off state.

Here we introduce and discuss experimental tests of a new feedback resistor model that postulates the presence of dissipative resistors in transactivation circuits and may explain off state stability in many animal virus circuits. We argue that feedback resistors can stabilize excitatory-feedback circuits by allowing them to generate a pulse of activity that eventually decays back to a stable off state (the only stable state). We used the HIV-1 Tat circuit as a model system, and we show, by measuring real-time expression kinetics at the single-cell level, that the circuit has a stable off state but is neither cooperative nor bi-stable. We then show that known enzymatic interconversions in the Tat feedback loop are sufficient to compose a feedback resistor and stabilize the Tat off state. The predicted pulses of Tat gene expression are directly observed in living single cells in real time, and biochemical and genetic perturbations of the suspected Tat resistor altered the stability of the Tat off state, in agreement with model predictions. The feedback resistor motif was more robust to noise than comparable cooperative-feedback motifs, and we postulate that this inherent noise buffering might allow feedback resistor circuits to explore a greater region of transcriptional space, i.e., different levels of transcriptional activity, while retaining off state stability. Desynchronized pulses in expression mediated by the feedback resistor can account for phenotypic diversity in isogenic populations and, specifically, for our previous observation of phenotypic bifurcation in the Tat feedback circuit [34]. Thus, feedback resistor motifs might be a common tactic used by viruses to maintain off state stability and buffer against transcriptional noise.

## Results

### Repressor Versus Resistor

To explain how animal viruses lacking bi-stable λ-like repressor motifs could maintain stable transactivator off states, we hypothesized that an excitatory-feedback loop might be stabilized if the transactivator underwent reversible modification to obtain its final functional form (Figure 1C). While the transactivator was in this intermediate (i.e., unmodified and nonfunctional) form, we assumed that it could decay (i.e., turn over by its half-life). In general, for a hypothetical transactivator, this modification scheme can have one or more nonfunctional intermediates. Importantly, functional modification of the transactivator has been reported for numerous animal viruses [35–38].

The intermediate, nontransactivating states are analogous to introducing a dissipative resistor (e.g., heat loss or nonadiabatic) into an electrical circuit. In this scheme, the resistor also functions as a time delay and has a mathematical formalism with a gamma-distributed time delay [39]. If the resistor is strong enough and the backward kinetic rates overcome the forward kinetic rates, then the resulting feedback resistor model generates a transient pulse of expression but is mono-stable, having a single steady state: the off state (Figure 1D; mathematical analysis in Figures S1 and S4).

Metaphorically, imagine the feedback circuit is a pinball machine where a canonical "spherical cow" (the pinball), with a finite half-life, completes its reaction by dropping through the flippers at the bottom of the machine. Each time a ball drops, two new balls are shot out by the "plunger" (this is the positive feedback). Because pinball machines are tilted, a ball anywhere on the table will eventually drop through the flippers at the bottom, resulting in two new balls, each of which will eventually be lost through the bottom thereby creating more balls on the table (i.e., the system is unstable). The number of lost balls relates to the speed of the transactivation reaction. If the pinball machine has a bumper area where balls are bounced back and forth between bumpers, balls can conceivably be trapped bouncing back and forth indefinitely until they decay by their half-life (sticky fly paper could also slow and dissipate balls). If the bumpers are of sufficient size, they can be dissipative; all balls will eventually be trapped in the bumpers and decay away before being lost through the flippers to initiate the positive-feedback response (i.e., the system is now stable at zero). Essentially, the bumpers generate a futile cycle (also known as an enzymatic interconversion between two covalent states) and act as a dissipative resistor in the system.

Theoretically, any enzymatic covalent interconversion (e.g., acetylation/deacetylation, phosphrylation/dephosphoylation, sumolation/desumolation, and neddylation/deneddylation) could act as a dissipative resistor in a feedback circuit. Equilibrium in the rates of transcription and translation (specifically, mRNA and protein synthesis versus decay) can also generate a "weak" resistor in auto-regulatory circuits. The weak resistor is akin to lengthening the pinball machine table so that balls have a greater probability of decaying before being lost through the bottom. But, for such weak resistors to stabilize the off state of a transactivation circuit, the half-life of the transactivator must be relatively short and the strength of the positive feedback must be relatively weak. Biologically, the opposite appears to be true: HSV-1 ICP0 has a life of >8 h [40], human CMV IE1 has a half-life of >24 h [35], and HIV Tat has a half-life of ~8 h [41]. Thus, for these viral transactivation circuits, a weak resistor appears to be insufficient and a dissipative feedback resistor appears necessary to stabilize the off state. In other systems with short-lived transactivator species, equilibrium in the rates of transcription and translation may indeed make up a feedback resistor.

In this feedback resistor scheme above, "all roads lead to Rome" (i.e., all transactivation trajectories eventually decay to the off state, the circuit is now an excitable system). Below we analyze the kinetics of the HIV-1 Tat circuit and present evidence for a feedback resistor stabilizing the Tat circuit's off state.

## HIV-1 Tat: A Viral Feedback Circuit That Is Not Bi-Stable, but Mono-Stable in the Off State

Cells infected with full-length HIV-1 or minimal derivatives of the virus that include the Tat circuit are known to establish an off state in cell culture [27,42], but previous studies did not explicitly test bi-stability in the Tat circuit or examine the stability of the transactivated state. The stability of the Tat transactivated state is important, because Tat is believed to constitute a genetic switch [25,43].

To simplify our analysis and verify that off state stability was indeed a property of the Tat positive-feedback loop and not another viral gene product, we restricted our study to minimal HIV-1 Tat circuits containing only the LTR promoter expressing a single mRNA. The RNA included an internal ribosome entry site (IRES) to allow expression of two proteins: Tat and a green fluorescent protein (GFP) reporter. Jurkat cell clones containing four minimal HIV-1 circuits were examined: (i) an LTR-Tat-IRES-GFP circuit (LGIT), (ii) the reversed LTR-GFP-IRES-Tat circuit (LTIG), (iii) a two-color LTR-RFP-IRES-Tat-GFP (LRITG) fusion circuit, and (iv) a Tat minus LTR-GFP system as a control. These Jurkat clones, originally sorted from a region of "dim" GFP fluorescence, were previously characterized for their transcriptional and transactivation activity [34], and they have known viral integration sites in intergenic regions of the human genome (thus basal expression from the LTR promoter is quite low and can be neglected). Previous work also verified that chromatin silencing mechanisms were not acting at the viral integration sites in these clones during extended periods of culture. Mathematically, these clones allowed us to restrict our study to circuits that conformed to simple feedback models of the form: $dTat/dt = basal + \Gamma[Tat(t)]$, where $basal \approx 0$, (and $\Gamma$ represents a generalized Tat transactivation function to be determined below).

The kinetic relaxation from transactivated to off for these Jurkat clones was measured by fluorescence activated cell sorting (FACS). Cells expressing high levels of GFP (and Tat) were isolated from different clones, and the proportion of cells that maintained GFP expression after further culture periods was assayed. The vast majority of transactivated GFP+ cells eventually relaxed into the off state (Figure 2A), thus indicating that the Tat circuit was mono-stable in the off state, as opposed to bi-stable (stochastic noise in the circuit appeared to be responsible for a continuous background level of activation and thus incomplete relaxation of all cells in the population, see below). Comparable relaxation into a mono-stable off state was observed in many parallel experiments, including: (i) infections with LTR-GFP-IRES-Tat, LTR-Tat-IRES-GFP, or LTR-mRFP-IRES-Tat-GFP viruses eventually relaxed into a GFP off state; (ii) GFP+ cells of each of these infections, isolated by bulk FACS, eventually relaxed into an off state; and (iii) GFP+ cells of each of these infections isolated by single-cell (i.e., clonal) FACS analysis eventually relaxed into an off state (unpublished data). The observed rate of relaxation into the off state (Figure 2A) is largely due

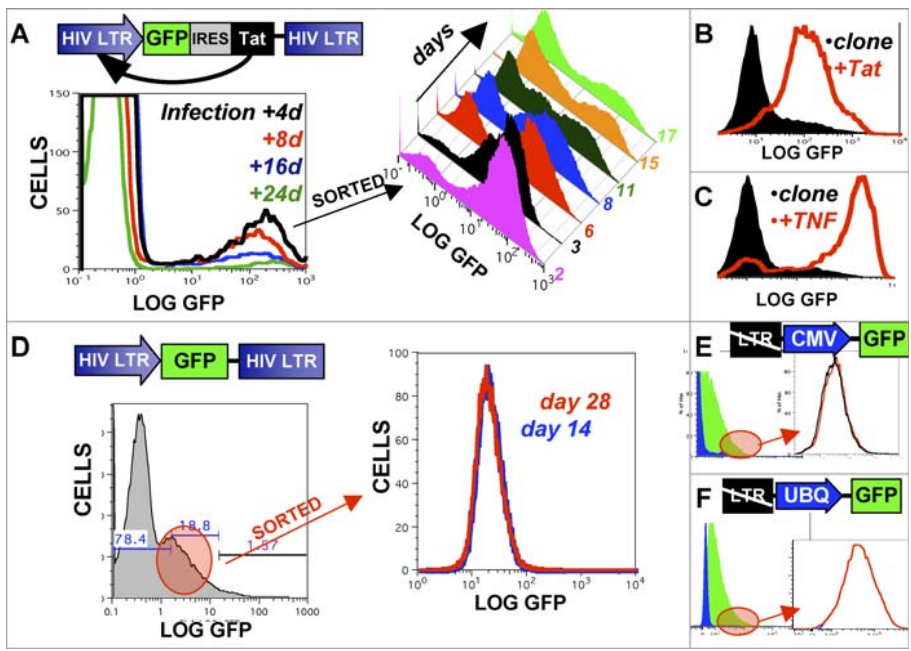

**Figure 2.** The HIV-1 Tat Circuit Is Not Bi-Stable

(A–C) Jurkat cells infected with an LTR-GFP-IRES-Tat virus (at a multiplicity of infection MOI ≈ 0.1) and analyzed by flow cytometry as the GFP+ population relaxed to the off state (left panel). GFP+ cells were also isolated by FACS, placed in culture, and analyzed by flow cytometry for residual GFP expression after various time periods (right panel). FACS indicates that the GFP+ cells relaxed into an off state over time. (GFP+ population relaxation was also observed for single-cell clonal FACS). Off state GFP− cells (black) could be reactivated by incubation in either exogenous Tat protein (B) or TNF-α (red) (C). Exogenous Tat protein does not overactivate the transactivated subpopulation (i.e. the transactviated mean is not brighter than in the original LTR-GFP-IRES-Tat infected cells), whereas TNF-α does overactivate the transactivated subpopulation (i.e., the transactviated mean is significantly brighter than in original LTR-GFP-IRES-Tat infected cells).

(D–F) Jurkat cells infected with a control virus lacking Tat and having GFP expressed from either the HIV LTR (D), CMV major immediate early promoter (E), or the ubiquitin (UBQ) promoter (F). These infected cells were sorted via FACS to isolate GFP+ cells (left panels) and analyzed by flow cytometry after 14 days (blue) or after 28 days (red) in culture. GFP+ cells did not relax to an off state in these controls.

doi:10.1371/journal.pbio.0050009.g002

to the long GFP half-life (~48 h), Jurkat cell doubling time (~24 h), and LTR-mRFP–IRES-Tat-GFP cells, which have a GFP fusion that is approximately four times less stable, showed approximately four times quicker shutdown (Figure S5).

Importantly, off state cells can be reactivated with either exogenous Tat protein (Figure 2B) or tumor necrosis factor alpha (TNF-α) (Figure 2C), a compound known to activate the HIV LTR through its binding sites for the transcription factor NF-κB [44]. The finding that exogenous Tat protein alone can activate our clones implies that RNAPII can access the LTR promoter and transcribe the RNA TAR stem loop. It also argues strongly against epigenetic chromatin silencing blocking RNAPII promoter binding/initiation and thereby establishing the off state in these cells [27,45]. TNF-α reactivation indicates that off state cells are transcriptionally viable and have not lost or mutated the viral genome. Control circuits assembled into lentiviral vectors containing GFP but lacking Tat did not exhibit relaxation to the off state (Figure 2D), and the same was true for GFP expressed under control of the CMV major immediate-early promoter (Figure 2E), the ubiquitin promoter (Figure 2F), or elongation factor-1a promoter (unpublished data).

As noted earlier, simple noncooperative models of Tat feedback, e.g., $d/dt(Tat) = k_{TR} \times Tat(t)^H$ or $d/dt(Tat) = [k_{TR} \times Tat(t)^H]/[k_M + Tat(t)^H] - \delta \times Tat(t)$, where the cooperative Hill coefficient $H = 1$, $k_{TR}$ is the transactivation rate, $k_M$ is the Michaelis constant and $\delta$ is the Tat decay rate) cannot account for Tat off state stability. Both models are indeed equivalent at low Tat concentrations and rapidly leave the off state. But, introducing a nonlinearity (i.e., $H > 1$, which implies cooperativity and oligomerization) can produce a bi-stabile circuit having a stable off state [46]. We next set out to test directly whether Tat feedback exhibited cooperativity.

## Tat Feedback Is Noncooperative

Three independent lines of study were followed to test for cooperativity in the Tat circuit. The first study examined the expression kinetics of Tat transactivation in single cells, because linear- and cooperative-feedback models predict very different rates of feedback.

TNF-α, which activates the HIV LTR, was used to "jump-start" Tat feedback in Jurkat clones. All clones contained known single integrations of the minimal LTR-GFP-IRES-Tat virus within intergenic domains; they exhibited very low basal LTR expression and were not prone to epigenetic chromatin silencing [34]. To quantify the Tat transactivation rate, GFP expression kinetics were followed in live single cells in real time by fluorescence imaging after TNF-α administration (Video S1). The resulting single-cell transactivation trajectories (Figure 3A and 3B) showed that Tat transactivation exhibited noncooperative kinetics in time. At early times up to ~8 h, the trajectories could be fit by either the linear ordinary differential equation (ODE) model $d/dt(Tat) = k_{TR} \times Tat(t)$ having the solution $Tat(t) = Tat_0 \times \exp(k_{TR} \times t)$ (Figure 3A, red line), or a quadratic/polynomial equation (e.g., $Tat(t) =$ $Tat_0 \times k_{TR} \times t^2$). At longer times after treatment with TNF-α, saturation in feedback was observed, and the trajectory fit the saturable ODE: $d/dt(Tat) = k_{TR} \times Tat(t)/[k_M + Tat(t)] - \delta \times Tat(t)$ (Figure 3A, black line). To verify that noncooperative activation kinetics were not merely an intrinsic property of TNF-α activation or the maturation/folding rate of the GFP fluorophore, we activated LTR-GFP controls (lacking any feedback mechanism) with TNF-α and LTR-GFP activation kinetics were linear in time (Figure 3C). Importantly, cooperativity leads to rates of increase that are greater than linear (nonlinear) on a log scale (Figure 3D), but single-cell kinetics showed a linear (or slightly sublinear) increase in GFP expression on a log scale (Figure 3E). The slight deviation from linearity (slight sublinearity) maybe due to either transactivation saturation (Figure 3A), GFP decay, or a quadratic feedback process. To verify that noncooperative kinetics resulted from perturbations of the LTR, we analyzed the kinetics of GFP expression after feedback circumvention using exogenous Tat protein and found these GFP kinetics to be subexponential in time (Figure S2). The decay rate of a Tat-GFP fusion protein was also measured using this single-cell technique, and the Tat-GFP half-life after cycloheximide addition was measured at >8 h, in accordance with published values [47] (unpublished data).

Thus, this single-cell system appears to measure accurately transcriptional kinetics and shows that expression from the Tat circuit is noncooperative. Furthermore, Tat transactivation kinetics cannot be explained by a nonlinear feedback model, thereby arguing against cooperativity and supporting the finding above that Tat transactivation is not bi-stable (bistability requires a nonlinearity in the feedback model).

To assay directly for Tat cooperativity, the Hill coefficient for transactivation was quantified by measuring the dose response of Tat on the HIV LTR. The Hill coefficient *(H)* describes the degree of enzymatic self-cooperativity, with $H=1$ implying no cooperativity (as in the case of linear feedback) and $H > 1$ signifying cooperativity in an enzymatic reaction (i.e., enzyme dimers or higher-order multimers are required for catalysis). We assumed that Tat acted as an enzyme on the HIV LTR (which is not limiting) and produced the GFP product such that: $GFP(Tat) = \frac{V_{max}Tat^H}{k_M + Tat^H}$. We exposed a well-characterized LTR-GFP Jurkat clone [34] to increasing amounts of Tat protein and measured the LTR GFP activity by flow cytometry (Figure 4A). Nonlinear least-squares regression of the data robustly generated a Hill coefficient $H = 1 \pm 0.1$, and the fit error was determined by re-fitting the data after repeated removal of a random data point (i.e., jack-knifing) and by varying the initial input guesses for regression between $H = 0$ and $10$. As a further test of the robustness of the fit, we forced $H = 2$ and repeated the regression, but very poor data fits were generated (unpublished data). Thus, Tat dose-response data, which shows no Tat self-cooperativity, agrees precisely with single-cell Tat transactivation kinetics showing non-cooperative feedback. As a further test of the noncooperativity of Tat feedback, we verified the prediction that the Tat dose-response curve for the full feedback circuit LTR-GFP-IRES-Tat should be linear (because $\partial(k_{TR}Tat^H)/\partial Tat = k_{TR}$ when $H = 1$) (Figure S3).

Finally, because oligomerization is typically required for cooperativity, we assayed directly for Tat oligomerization in vivo using a fluorescence resonance energy transfer (FRET) technique that can differentiate between monomers, homo-

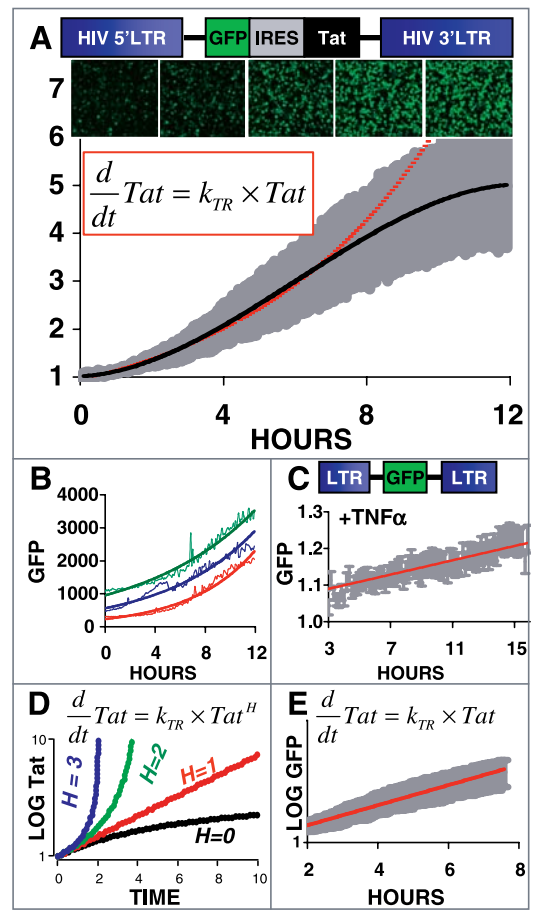

**Figure 3. Single-Cell Tat Transactivation Kinetics Are Noncooperative**

(A) Single-cell kinetics of LTR-GFP-IRES-Tat Jurkat cells after activation with TNF-α. Fluorescence intensities for 35 single cells over time were normalized (gray). At times <8 h, the average intensity fit the noncooperative minimal feedback model: $d/dt(Tat) = k \times Tat$ or its exponential solution $Tat(t) = Tat_0 \times exp^{kt}$ (red line). A quadratic expression (also noncooperative) provides an equally good data fit at early times, and it is difficult to distinguish between exponential and quadratic (unpublished data). At later times after TNF-α treatment, GFP fluorescence plateaus, possibly due to the intrinsic dynamics of the NF-κB response [70], but transactivation kinetics still fit the noncooperative saturating feedback model: $dTat/dt = (k_{TR} \times Tat)/(k_M + Tat) - \delta \times Tat$ with $k_{TR} \approx 8$, $k_M \approx 0.08$, and $\delta \approx 2$ as determined by nonlinear least-squares regression. Since the minimal model $dTat/dt = k_M \times Tat$ fits the data essentially perfectly for the first 8 h, indicating that Michaelis-Menten saturation is unnecessary for fitting short-term <8-h data, for simplicity we focus on the minimal model and <8-h data.

(B) Representative non-normalized, single-cell data used to construct panel (A).

(C) Control: LTR-GFP kinetics after TNF-α activation are linear in time, indicating that the exponential (or quadratic) Tat activation kinetics cannot be explained by an intrinsic property of TNF-α activation.

(D) Log scale plots of the solution to $d/dt(Tat) = k_{TR} \times Tat^H$ for increasing values of cooperativity (i.e., the Hill coefficient, $H$). Both the cases where $H = 0$ and $H = 1$ are noncooperative, and the $H = 1$ case is linear on a log scale.

(E) LTR-GFP-IRES-Tat transactivation kinetics [from (A)]) replotted on a log scale, showing that Tat transactivation kinetics more closely match a noncooperative system (i.e., $H \approx 1$).

doi:10.1371/journal.pbio.0050009.g003

dimers, and higher order homo-multimers [48]. This technique, called homo-FRET, uses GFP's inherent polarization anisotropy *(r)* to measure homo-oligomerization of a GFP fusion protein (in this case, Tat-GFP driven off the HIV LTR). GFP monomers freely diffusing in solution or in the

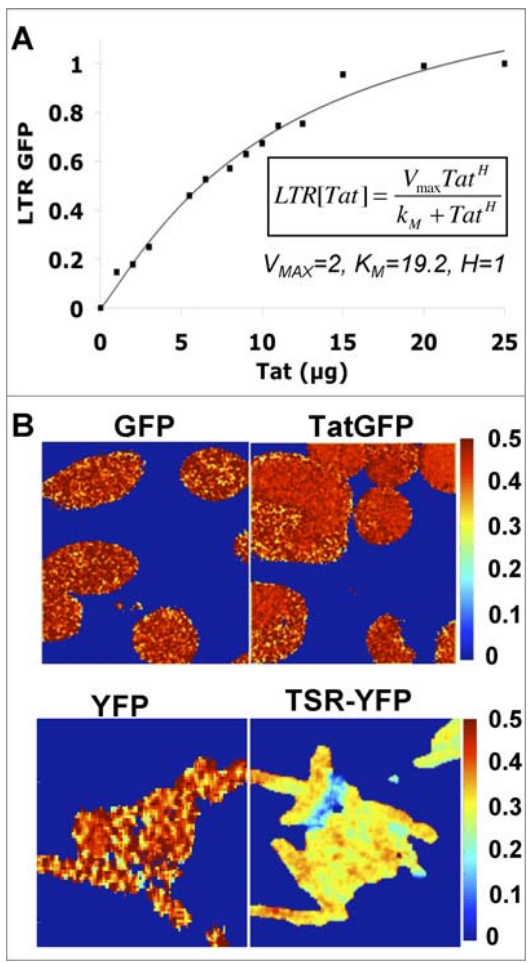

**Figure 4.** Tat Functions in a Noncooperative Manner

(A) An LTR-GFP clone was exposed to varying amounts of purified Tat protein and analyzed by flow cytometry 12 h later. Data were fit by nonlinear least-squares regression to a Michaelis-Menten model (inset) allowing the cooperative Hill coefficient to vary. The Hill coefficient was robustly measured to be 1 under all initializations tested.

(B) Polarization anisotropy (homo-FRET) analysis of Tat-GFP in single Jurkat cells infected with either LTR-GFP or LTR-IRES-TatGFP virus. A TSR-YFP fusion protein known to multimerize and exhibit homo-FRET exchange [69] was used as a positive control. Polarization anisotropy $r$ values are shown in color with color bar at right. Cells expressing either (monomeric) GFP or YFP displayed polarization $r$ values of $r \approx 0.47$ near the theoretical two-photon limit of $r = 0.5$ (red). Cells expressing TSR-YFP displayed significantly reduced polarization $r$ values ($r \approx 0.3$, yellow), indicating homo-FRET exchange. Cells expressing Tat-GFP displayed a slight increase in polarization anisotropy ($r \approx 0.5$), indicating that no homo-FRET was occurring.

doi:10.1371/journal.pbio.0050009.g004

cellular cytoplasm have a known two-photon polarization anisotropy of $r \approx 0.5$ [49]. Fusing a large protein to GFP alters its diffusion and typically increases $r$ ($>0.5$) unless the fusion protein molecules are in close proximity, in which case $r$ is decreased ($<0.5$). A two-photon microscope to measure polarization anisotropy was constructed and homo-FRET exchange was verified for a previously reported multimerizing protein, the *Escherichia coli* membrane-spanning serine chemoreceptor TSR. A TSR-YFP fusion protein exhibited $r$ values that were significantly lower than free yellow fluorescent protein (YFP) in cells (Figure 4B, bottom), indicating that homo-FRET exchange had reduced the polarization anisotropy. GFP expressed from the HIV LTR in Jurkat cells exhibited polarization anisotropy values of $r \approx 0.5$, as expected, whereas Tat-GFP expressed from the HIV LTR in Jurkat cells exhibited no decrease in polarization anisotropy (in fact, a slight increase was observed; Figure 4B, top right). This is consistent with Tat being a monomer [31,32] and with the observed lack of self-cooperativity in the HIV Tat feedback loop.

In summary, the data discussed so far (no Tat bi-stability, noncooperative Tat transactivation kinetics, Tat Hill coefficient = 1, and monomeric Tat) are parsimonious and support a model where Tat feedback is noncooperative. Had Tat exhibited oligomerization, it would have likely exhibited self-cooperative feedback and would have produced a bi-stable circuit, where both the on state (transactivated state) and the off state would be stable. But, results throughout this study (including results in Figure 2 and those discussed below) show that the Tat off state is the only stable state, again arguing against bi-stability and cooperativity.

Given the lack of cooperativity in the Tat transactivation circuit, the feedback resistor model appears to be a necessary condition for off state stability in the Tat circuit.

## A Noncooperative Feedback Resistor Can Account for Off State Stability

Here we present a minimal model that can explain Tat off state stability and observed transactivation kinetics. The model is presented in general form with the molecular details highly simplified, and it represents the simplest model that we believe can explain transactivation kinetics and Tat off state stability. Importantly, the model does not preclude other molecular mechanisms; it merely presents the simplest necessary conditions for the observed transactivation kinetics and off state stability.

The model focuses on transactivation involving two distinct steps [24,37,50] and enzymatic interconversion between the two forms of Tat that mediate progression through the transactivation steps. First, deacetylated Tat ($Tat_D$) binds to the nascent TAR RNA loop in the LTR and recruits pTEFb, which phosphorylates RNAPII thereby rendering it highly processive for elongation. But, the nuc-1 is believed to be sterically blocking the elongation of RNAPII. Once $Tat_D$ is converted into acetylated Tat ($Tat_A$) by the host histone acetylatranferase p300, $Tat_A$ recruits the host SWI/SNF chromatin-remodeling complex to the LTR, thereby facilitating remodeling of nuc-1 and allowing the highly processive RNAPII to proceed downstream [24]. Clearly, $Tat_A$ can be considered the rate-limiting step for transactivation and completion of the feedback loop. $Tat_A$ is also deacetylated by the host histone deacetylase SirT1, and Tat deacetylation occurs at a much faster rate than Tat acetylation; immediately after microinjection of synthetic $Tat_A$ into cells, only $Tat_D$ can be detected [25], indicating that acetylated Tat is almost immediately converted to deacetylated Tat. Thus, the enzymatic interconversion between $Tat_A$ and $Tat_D$ (along with the accompanying SWI/SNF chromatin remodeling) constitutes the minimal requirement for a feedback resistor. While NF-$\kappa$B p50-HDAC1 complexes can repress the TAR RNA binding target via chromatin restriction on transcriptional initiation, this appears to occur primarily for integrations in highly compacted alphoid-satellite repeats [45]. A schematic of the HIV-1 Tat feedback resistor model is

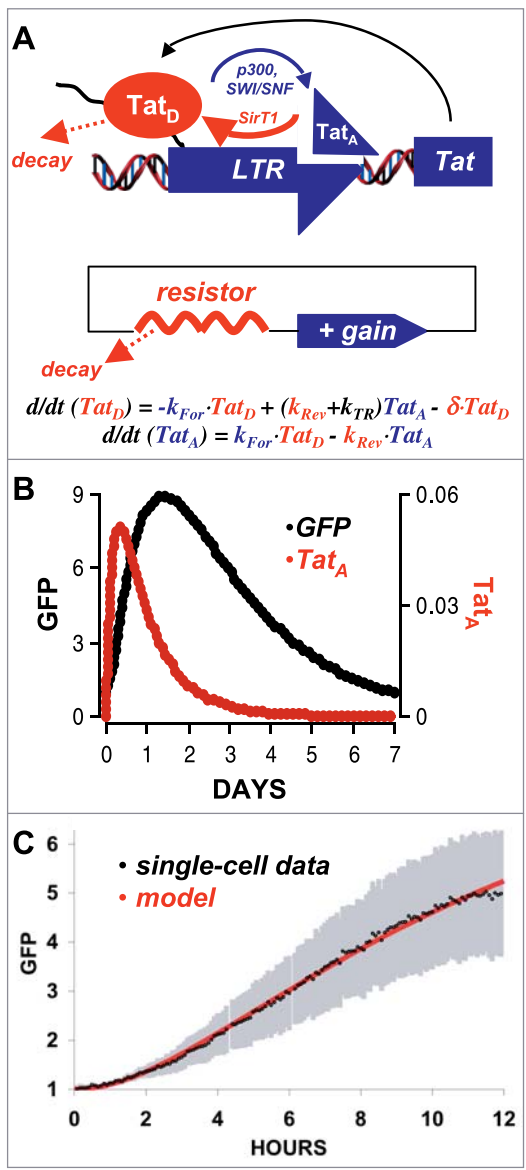

**Figure 5.** The Feedback Resistor Model

(A) A simplified schematic of the feedback resistor model for HIV-1 Tat transactivation along with the analogous circuit diagram and color-coded differential equations describing the system. Deacetylated Tat protein ($Tat_D$, red oval) is in excess in the cell and, along with CDK9 and CyclinT1 (not shown), binds the TAR RNA loop (black line) at the HIV-1 LTR and is acetylated by p300 (blue arrow). Acetylated Tat protein ($Tat_A$, blue triangle) is the limiting reagent, which completes the transactivation loop by recruiting SWI/SNF. However, SirT1 deacetylation (red arrow) can back-convert $Tat_A$ to $Tat_D$, and SirT1 deacetylation of $Tat_A$ is significantly faster than p300 acetylation of $Tat_D$. To account for intrinsic dynamics of the NF-κB response and NF-κB p50-HDAC effects on local chromatin environment, a time-varying basal expression parameter ($k_{basal}$) can be included in the $Tat_D$ equation, but this was unnecessary and did not qualitatively alter the behavior of the model.

(B) Numerical simulations of Equations 1 and 2 plus an added reporter equation, which has standard Michaelis-Menten promoter saturation, for tracking GFP expression. $d/dt(GFP) = IRES \cdot k_{TR} \cdot Tat_A/(k_M + Tat_A) - \delta_{GFP} \cdot GFP$. Simulations show that the feedback resistor model produces a stable off state and generates a pulse of transactivation over several days [GFP shown in black, $Tat_A$ shown in red; parameters used: $k_{for} = 0.5/d$, $k_{rev} = 5/d$, $k_{TR} = 5/d$, $\delta_{Tat} = 2/d$, $\delta_{GFP} = 0.5/d$, IRES = 50, $k_M = Tat_D(0) = 0.001$, $GFP(0) = 1$].

(C) Nonlinear least-squares regression of the model to the single-cell kinetic data from Figure 3A. The initial rise in the pulse of GFP expression (red line) predicted by the simulation in (B) matches the single-cell data from Figure 3A (black circles are the mean). Incorporating Michaelis-Menten saturation effects into other processes/parameters in the model did not qualitatively change the model behavior or the fit (unpublished data).

presented in Figure 5A, and a mathematical description of the model follows. To maintain generality and leave room for future molecular details that will likely be characterized, we used the commonly used "lumped" parameter method, where the rates $k_{TR}$, $k_{for}$, and $k_{rev}$ can represent numerous molecular processes that may be occurring.

### Tat Feedback Resistor Model

$$\frac{d}{dt}Tat_D = -k_{for}Tat_D + k_{rev}Tat_A + k_{TR}Tat_A - \delta Tat_D \quad (1)$$

$$\frac{d}{dt}Tat_A = k_{for}Tat_D - k_{rev}Tat_A \quad (2)$$

This minimal model accounts only for the populations of $Tat_A$ and $Tat_D$ protein. "Lumped" kinetic rates in the forward ($k_{for}$) and reverse ($k_{rev}$) directions can account for acetylation/deacetylation and other enzymatic/molecular interconversions. The model can be explained by the following reaction scheme: (i) $Tat_D$ (produced by translation) is converted to $Tat_A$ ($-k_{for}Tat_D$ in Equation 1 and $+k_{rev}Tat_A$ in Equation 2); (ii) $Tat_A$ is then rapidly back-converted ($-k_{rev}Tat_A$ in Equation 2) but can transactivate before being back-converted ($+k_{TR}Tat_A$ in Equation 1), thereby producing more $Tat_D$. p300-mediated acetylation of $Tat_D$ along with recruitment of SWI/SNF likely accounts for a large portion of the $k_{for}$ parameter, whereas deacetylation of $Tat_A$ by SirT1 likely accounts for a large component of the $k_{rev}$ parameter.

Importantly, this minimal model can also capture SirT1-mediated positive recycling of Tat as reported by Pagans et al. [50], because the term $k_{rev}Tat_A$ (Figure 5A) can represent both SirT1 back-conversion of acetylated Tat (before transactivation completion) and recycling of acetylated Tat (after transactivation). In Equations 1 and 2, $Tat_D$ does not transactivate the LTR but retains the capacity to do so if it is acetylated (and recruits SWI/SNF). Thus, $Tat_D$ is the dissipative feedback resistor: the $Tat_D$ compartment is a reservoir from which Tat decays but feeds into $Tat_A$ relatively slowly, thereby delaying and dissipating transactivation progression.

The decay of $Tat_D$ is modeled as exponential decay in accord with the reported half-life [41], but the decay of acetylated Tat is neglected, because it presents only in a short-lived multi-protein complex at the HIV LTR. Basal transcription is also assumed to be negligible in the clones we examined (see above; for analytic solutions to Equations 1 and 2), but NF-κB p50-HDAC1 complex chromatin repression of transcriptional initiation and TAR RNA [45] can easily be included in a $k_{basal}$ parameter. In general, the feedback resistor model does not preclude the addition of a basal expression term. Steady-state stability analysis (Figure S1) shows that the feedback resistor model produces a stable off state (eigenvalues < 0) when the following stability criterion is satisfied:

### Off State Stability Criterion

$$(k_{for}k_{TR}) < (k_{rev}\delta) \quad \textbf{ST1}$$

Intuitively, the stability criterion **ST1** is satisfied (thereby stabilizing the off state) when backwards reaction rates overcome forward reaction rates. In agreement with this criterion, the rate of Tat deacetylation is known to be much faster than the rate of acetylation [37], and the necessary recruitment of SWI/SNF only enhances this difference. Simulations of the Tat feedback resistor model confirmed that the off state was stable and attracting under physiological parameter conditions but that small perturbations could drive a pulse of LTR activity (Figure 5B). Furthermore, incorporating a GFP reporter into the model allowed us to compare directly the model to the single-cell kinetic expression data. Simulations of GFP activation after TNF-α activation of LTR-GFP-IRES-Tat or LTR-Tat-GFP showed an increase in GFP over time that agreed with the single-cell experimental results (Figure 5C). Introducing a time delay ($+k_{TR} Tat_A(t - \tau)$ in Equation 1) has no affect on the stability of the off state, but can change the shape of the transactivation pulse (unpublished data).

## Direct Visualization of the Predicted Transcriptional Pulse

To test the specific prediction that Tat generates a pulse of transactivation, LTR-GFP-IRES-Tat and LTR-GFP clones were "jump-started" using exogenous Tat protein (a form of Tat that is deacetylated). LTR-GFP-IRES-Tat clones transactivated with exogenous Tat protein exhibited a pulse of transactivation lasting several days, and the transactivated population relaxed back to the off state over approximately 2 wk (Figure 6A). A transactivation pulse was also observed in single cells activated with exogenous Tat protein (Figure 6B, black squares). This pulse had extremely rapid kinetics, with decay of transactivation evident at 9 h after the addition of Tat. This cycle is far faster than could be predicted from the >8-h Tat protein half-life alone (Figure 6B, red line) [41] and could be due to rapid Tat deacetylation. To test this idea, the experiment was repeated using cells in which SirT1 is overexpressed from a retrovirus vector [51] (Figure 6B, orange line). SirT1 deacetylates Tat [50], and its overexpression decreased the upward slope of the pulse. Thus, the kinetics of the transactivation pulse lends further support to the view that Tat deacetylation accounts for a significant component of the feedback resistor, and suggested experiments to perturb the resistor.

## Perturbing the Tat Feedback Resistor Alters Off State Stability and Single-Cell Activation Kinetics

To test whether the feedback resistor controlled Tat off state stability, we perturbed the kinetic parameters $k_{for}$ and $k_{rev}$ to change the stability relation **ST1**. We hypothesized that p300 and SirtT1 activities were significant components of the $k_{for}$ and $k_{rev}$ parameters, respectively.

Specifically, decreasing $k_{rev}$ (by inhibiting SirT1 activity) should mitigate the strength of the feedback resistor and make the stability relation **ST1** less likely to be satisfied, thereby destabilizing the off state. LTR-GFP-IRES-Tat Jurkat clones were exposed to several SirT1 inhibitors (nicotinamide, sirtinol, splitomycin, HR73, and dihydrocoumarin), and the occupancy of the off state was measured by flow cytometry. All SirT1-inhibiting drugs destabilized the off state and produced a marked induction of transactivation (Figure 7A, nicotinamide red, upper panel, other data not shown). In contrast, SirT1 inhibitors had no affect on control

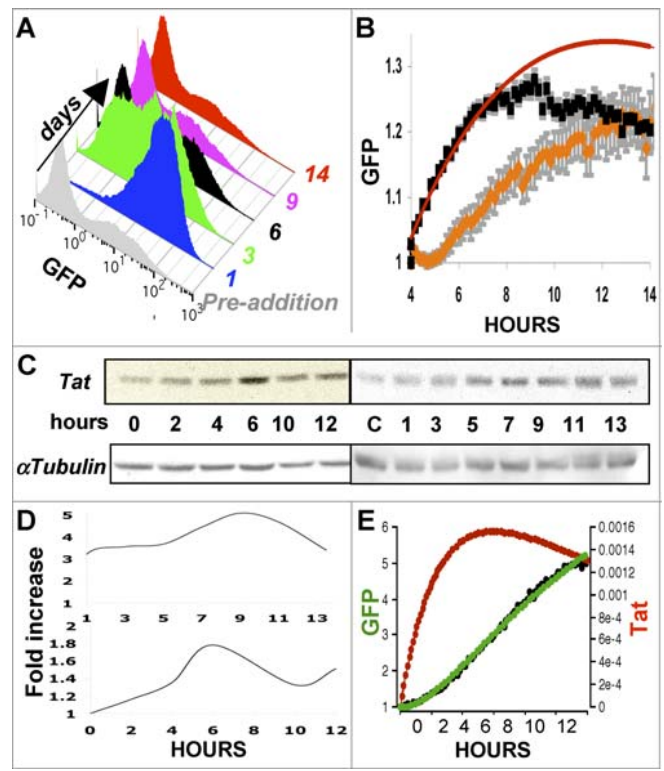

**Figure 6.** Direct Visualization of the Feedback Pulse

(A) An LTR-GFP-IRES-Tat Jurkat clone was incubated in medium containing exogenous Tat protein for 4 h, and then GFP expression was followed by flow cytometry for 2 wk. Transactivated cells eventually relaxed back into the off state. In a simple positive-feedback circuit, transactivated cells would be predicted to remain GFP$^+$.

(B) Pooled single-cell trajectories (average of ~100 cells) of an LTR-GFP Jurkat clone incubated in medium containing Tat protein for 4 h and then jumped (black squares) for 10 h. The downward trajectory of the GFP pulse begins at 9 h after addition of exogenous Tat, far quicker than predicted in a non–feedback resistor simulation based on a Tat half-life of 8 h (red line). The transactivation decay was based on Tat half-life alone (red line). Performing the same exogenous Tat incubation with cells in which SirT1 is overexpressed from a retrovirus vector decreases the upward slope of the pulse (orange). Error bars are shown as gray background behind the data points.

(C) Western blot analysis of Tat transactivation kinetics in LTR-GFP-IRES-Tat cells after exposure to TNF-α. Analyses were split into even and odd time points. The Tat signal band was visible at ~15 kDa, whereas the control α-tubulin signal was visible at ~50 kDa. The Tat protein level shows a pulse, peaking at 6–8 h after TNF-α exposure, despite constant α-tubulin levels.

(D) Quantitative densitometry analysis of the Western blots in (C). Tat data was normalized by subtraction of α-tubulin data for corresponding time points and plotted as fold increase relative to the first time point.

(E) A replotting of regression fit from Figure 5C but with the simulation for Tat included, showing that the model predicts both an exponential/quadratic increase in GFP and a pulse of Tat$_A$ with a peak at 6–8 h.

doi:10.1371/journal.pbio.0050009.g006

LTR-GFP expression (unpublished data). Resveratrol is an activator of SirT1 [52] and was hypothesized to increase $k_{rev}$ and thus re-stabilize the off state. As predicted, resveratrol mitigated the nicotinamide-mediated increase in transactivation, re-stabilizing the off state (Figure 7A, upper panel, green). Although it is possible that SirT1 inhibitors are acting at many levels, the activation level produced by SirT1 inhibitors is comparable to the activation produced by exogenous Tat protein incubation (Figure 7A, lower panel), which does not alleviate repressive chromatin or enhance transcriptional initiation. Furthermore, TSA and TNF-α,

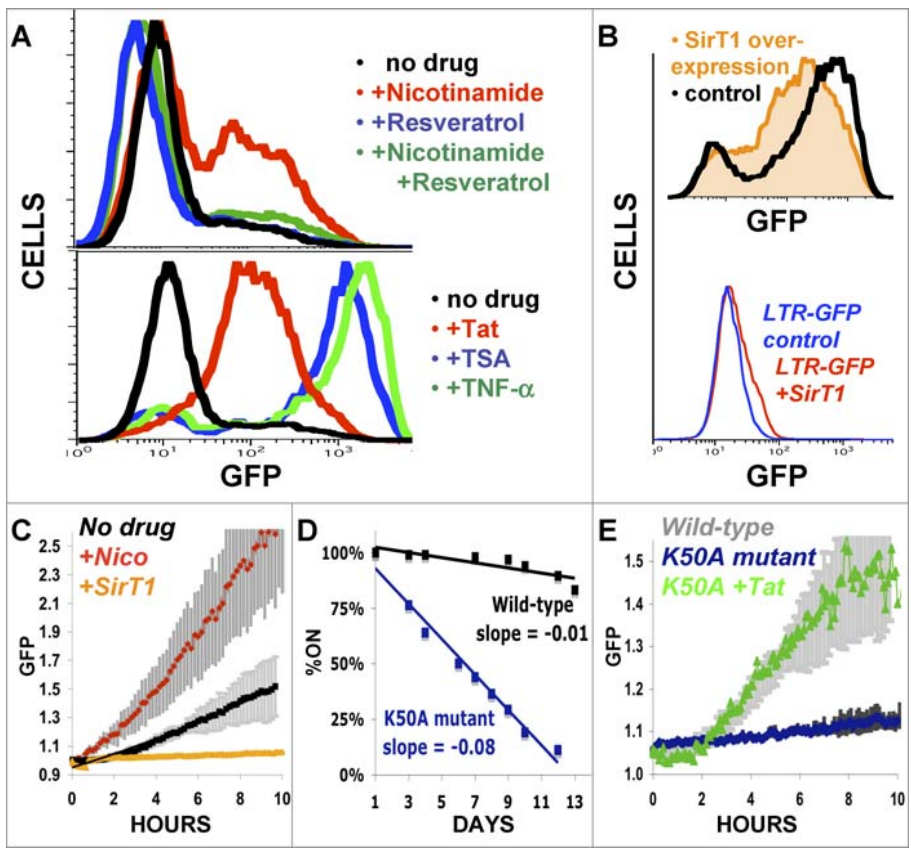

**Figure 7.** Perturbation of the Feedback Resistor and Off State Stability

(A) SirT1 inhibitors diminish the feedback resistor and destabilize the off state, but the effects are mitigated by co-incubation with the SirT1 activator resveratrol, which restabilizes the off state. An LTR-GFP-IRES-Tat Jurkat clone (E7) was exposed to either no drug (upper panel, black), nicotinamide (upper panel, red), resveratrol (upper panel, blue), or the combination of nicotinamide plus resveratrol (upper panel, green) for 72 h, and GFP expression was quantified by FACS. Off state destabilization via SirT1 inhibition is comparable to activation by exogenous Tat protein (lower panel, red), which does not enhance transcriptional initiation or alter local chromatin state. However, SirT1 inhibition is not comparable to activation by TSA (lower panel, blue) or TNF-α (lower panel, green), which do enhance transcriptional initiation by altering local chromatin state, and result in significantly increased activation.

(B) SirT1 overexpression amplifies the feedback resistor and stabilizes the off state. Upper panel: a Jurkat clone (LTR-GFP-IRES-Tat) was infected with a neomycin-resistant retrovirus expressing either SirT1 (orange) or an *E. coli* coding region as a control (black). After selection of neomycin-resistant cells, GFP expression was monitored by FACS. Lower panel: same as upper panel except that an LTR-GFP Jurkat clone was infected with the SirT1 (red) or control (blue) overexpression vectors.

(C) Single-cell transactivation kinetics show that SirT1 inhibitors increase the rate of Tat-mediated activation. An LTR-GFP-IRES-Tat Jurkat clone was exposed to TNF-α (black), or TNF-α plus nicotinamide (red) and a subset of cells exhibited significantly faster transactivation kinetics in presence of nicotinamide. SirT1-overexpressing LTR-GFP-IRES-Tat cells were also exposed to TNF-α (orange). Trajectories shown are averages of ~300 cells, gray backgrounds represent error bars.

(D) A Jurkat clone expressing mutant Tat (LTR-GFP-IRES-Tat[K50A]) decays into the off state approximately 7-fold more quickly than a clone expressing wild-type LTR-GFP-IRES-Tat cells.

(E) Tat acetylation mutant LTR-GFP-IRES-Tat(K50A) has approximately 6-fold slower transactivation kinetics (black) as compared to LTR-GFP-IRES-Tat (gray), and exogenous Tat (green) can rescue the mutant producing wild-type transactivation kinetics.

doi:10.1371/journal.pbio.0050009.g007

which do alleviate repressive chromatin structures and enhance transcriptional initiation, greatly increase promoter activity and generate a much stronger activation that differs markedly from the effects of SirT1 inhibitors (Figure 7A, lower panel). Thus, these results argue that SirT1 inhibitors destabilize the off state by enhancing transcriptional elongation (at a fixed level of transcriptional initiation), similar to exogenous Tat activation, rather than alleviating repressive chromatin to enhance transcriptional initiation and increase promoter activity. This argument is further supported by the fact that SirT1 inhibitors do not activate LTR-GFP control cells (a system lacking Tat).

We also examined the effect of increasing $k_{for}$ by using the retroviral SirT1 overexpression system. In agreement with the drug results, retroviral-mediated SirT1 overexpression produced a marked decrease in transactivation activity (mean GFP fluorescence) in LTR-GFP-IRES-Tat Jurkat cells as compared to a control infection, whereas SirT1 overexpression had no affect on LTR-GFP cells (Figure 7B). These results are consistent with the postulated role for SirT1 in the feedback resistor and support the accuracy of the stability relation **ST1**.

Perturbation of the kinetic parameters should also lead to a corresponding amplification or reduction in the feedback resistor and thus the transactivation kinetics. To quantify this change, real-time single-cell transactivation kinetics were measured after treatment of LTR-GFP-IRES-Tat cells with TNF-α. Multiple SirT1 inhibitors increased the Tat trans-activation rate, whereas SirT1 overexpression decreased the Tat transactivation rate (Figure 7C and unpublished data).

Next, the effect of decreasing the $k_{for}$ parameter was examined. Decreasing $k_{for}$ should enhance the stability criterion **ST1** and should thus increase off state stability. The $k_{for}$ parameter was perturbed using a previously characterized Tat K50A mutant [53] with a K→A substitution at position 50 (the major residue acetylated by p300). Jurkats were infected with this LTR-GFP-IRES-Tat(K50A) virus, and clones exhibiting low basal GFP expression were isolated and found to exhibit transactivated GFP fluorescence levels comparable to wild type but off state stability was increased compared to wild-type LTR-GFP-IRES-Tat clones, in agreement with our previous observations [34]. However, in the present study, we were able to demonstrate that a mutant LTR-GFP-IRES-Tat(K50A) clone exhibiting transactivated GFP fluorescence levels comparable to wild type relaxed to the off state approximately 7-fold more quickly than a wild-type LTR-GFP-IRES-Tat clone (Figure 7D), and we verified that this Tat(K50A) mutant clone also exhibited approximately 7-fold slower transactivation kinetics (Figure 7E). Importantly, slower Tat(K50A) kinetics were not a result of a secondary mutation in the LTR or elsewhere, because wild-type kinetics could be re-established upon the addition of exogenous Tat protein (Figure 7E). Thus, single-cell transactivation kinetics support the Tat feedback resistor model.

## The Feedback Resistor Buffers against Stochastic Noise in Gene Expression

Having determined that the Tat feedback circuit is not nonlinear/cooperative or bi-stable, and having proposed and tested an alternate feedback resistor model to explain the off state stability, we next explored the merits of the feedback resistor versus cooperative feedback. Our goal was to find a fitness criterion that might explain why a feedback resistor may have been selected over a cooperative-feedback mechanism.

We built upon our previous finding that the Tat circuit is subject to significant stochastic fluctuations [34], and we hypothesized that the feedback resistor might act as a buffer to establish a more stable off state than cooperative feedback. To test this hypothesis computationally, we first explored the effect of idealized noise, i.e., Langevin noise [54], on the simplest cooperative- and noncooperative-feedback equations (Figure 8A). Numerical simulations show that as cooperativity is increased (i.e., the Hill coefficient is increased above $H = 1$) a previously stable feedback circuit becomes unstable. Thus, cooperativity destabilized the off state of a simple feedback circuit in a noisy environment, and noncooperative feedback (e.g., feedback resistor model) appeared to be more robust to noise. These results agree with our previous Monte-Carlo simulation results for the HIV-1 Tat circuit [34] as well as other groups' findings [4,5]. An explanation for cooperativity's destabilizing effect is presented below in the Discussion section.

We next explored whether the feedback resistor model (Equations 1 and 2), which is noncooperative, also displayed robustness to stochastic molecular fluctuations (i.e., nonidealized, non-Langevin noise) and whether changing the strength of the SirT1 resistor altered the noise buffering capacity of the Tat circuit. Equations 1 and 2 were converted into a stochastic simulation [55], and Monte-Carlo computer

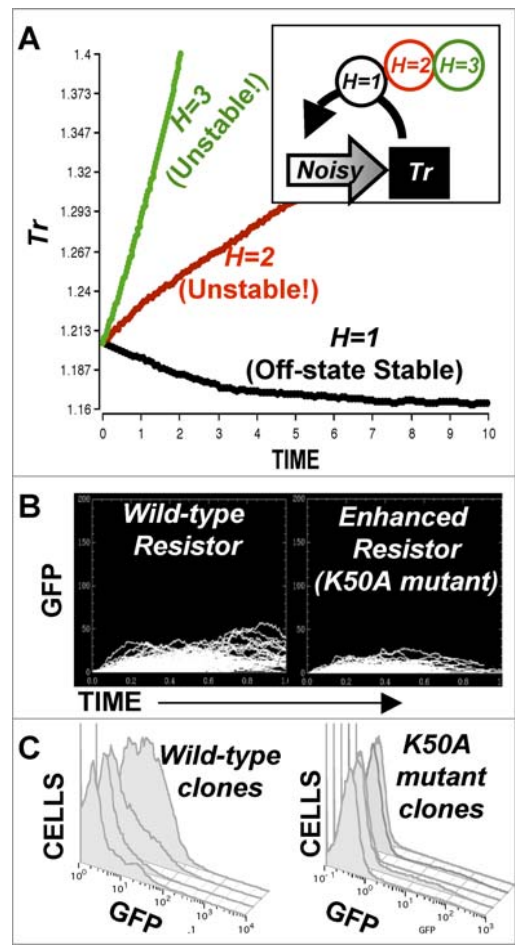

**Figure 8.** The Feedback Resistor is Relatively Robust to Molecular Noise

(A) Numerical simulations of the ODE: $dTr/dt = k_{TR} \times Tr^H/(k_M + Tr^H) - \delta \times Tr + NOISE(X)$, with a hypothetical transactivator (Tr) and increasing levels of cooperativity (H): Tr requires either monomers (H = 1), dimers (H = 2), or trimers (H = 3) to complete transactivation. The cooperative-feedback model (i.e., when H > 1) is more sensitive to noise than the H = 1 noncooperative feedback model (the feedback resistor is a noncooperative feedback model). All parameters (except H) were kept constant for all simulations: $k_{TR} = 1.6$, $k_m = 10$, $\delta = 0.5$, $NOISE(X)$ is a random number generated from a uniform distribution $X = [0, 1.6]$, and initial condition $Tr_0 = NOISE(0,2)$. Only H was varied between the simulations: $H = 1, 2,$ or 3. Similar results were also obtained with a simpler nonsaturating feedback ODE model (unpublished data).

(B) Simulation of a stochastic version of the feedback resistor model diagrammed in Figure 5A. Trajectories are direct Monte-Carlo simulations of the chemical master equation for Equation 1 and 2. Simulations predict that HIV-1 Tat levels continually fluctuate above the off state and that computationally strengthening the feedback resistor (by reducing the value of $k_{for}$ by 2-fold) decreases the level of the fluctuations above the off state.

(C) To test this prediction experimentally, single cells in the off state (GFP) were sorted with FACS from LTR-GFP-IRES-Tat and LTR-GFP-IRES-Tat(K50A) clones, grown for approximately 3 wk, and then analyzed by FACS for GFP expression. Off state fluctuations in the mutant clone were highly diminished compared to wild type, as predicted by the simulations. Four representative clones of cells containing wild-type and mutant Tat are shown from a total of 12 that were analyzed.
doi:10.1371/journal.pbio.0050009.g008

simulations were performed (Figure 8B). These simulations showed that the feedback resistor model maintained a stable off state (i.e., the trajectories did not transactivate), but the simulations also predicted that LTR-GFP-IRES-Tat cells populating the off state were in fact continually fluctuating slightly above the off state (Figure 8B, left panel). The simulations also predicted that these transcriptional fluctuations above the off state would be highly diminished if the feedback resistor was strengthened, as in the Tat(K50A) mutant (Figure 8B, right panel). To test these predictions experimentally, single cells were isolated by FACS from the off state of wild-type LTR-GFP-IRES-Tat and the mutant LTR-GFP-IRES-Tat(K50A) Jurkat clones described in Figure 7D. After an extended period of growth, cells sorted from the wild-type clone exhibited significant GFP fluctuation above the off state (Figure 8C, left), whereas cells sorted from the mutant K50A clone exhibited significantly less leakage from the off state. This result suggests that stochastic non-synchronized pulses in Tat transactivation likely account for our previous observation of phenotypic bifurcation in isogenic Tat circuit populations [34]. At any given cross-section of time after activation of an isogenic population, a subpopulation of cells will be at the peak of the transactivation pulse, while the remainder of cells will be in various states of decay from this peak with the majority accumulating in the off state.

## Discussion

Here we propose a feedback resistor mechanism to explain how excitatory, positive-feedback (i.e., transactivation) circuits maintain a stable transcriptional off state. Enzymatic interconversion of the transactivator molecule (a futile cycle) generates intermediate molecular species that make up a feedback resistor and allow the transactivator to decay by its half-life before it can positively feed back. Previously, the λ-phage repressor circuit, a highly cooperative transcriptional regulatory circuit, was the primary model used to explain bi-stable genetic circuits and viral latency. But many animal virus regulatory circuits apparently lack λ-like repressors, and we examined one of these excitatory viral feedback circuits (the HIV-1 Tat circuit) as a model system to explore and test the feedback resistor model. Real-time single-cell transcriptional kinetics showed that HIV-1 Tat feedback was neither bi-stable nor cooperative (Figures 2–4). The feedback resistor was then shown to stabilize the HIV-1 Tat off state by introducing an intermediate resistor motif into the feedback circuitry (Figure 5), and enzymatic interconversion between the acetylated and deacetylated forms of Tat appears to make up a significant component of the feedback resistor. The feedback resistor model predicted a pulse of transcriptional activity that eventually decayed to the off state, and we observed this counterintuitive prediction in the Tat feedback circuit directly (Figure 6). A pulse of transactivation is likely sufficient to initiate lytic replication and allow a virus to complete its intracellular life cycle, especially for HIV-1, which quickly kills infected cells [56].

In principle, a feedback resistor can help stabilize the off state or zero state of any transactivation circuit, even those exhibiting cooperativity, as has been reported to be the case for the Kaposi sarcoma–associated herpesvirus Rta transactivator [57]. Recent findings that the opposing actions of Rpd3 histone deacetylase and set2 histone methyltransferase buffer against spurious intragenic transcription [58,59] suggest that transcriptional repressors might also exploit the feedback resistor. All that is required for a resistor is that the backward (inhibitory) rates overcome the forward (excitatory) rates. Because transcriptional positive-feedback loops are common regulatory motifs found throughout signal transduction biology and in diverse viral families, resistor motifs might be common as well.

## Relative Merits of a Feedback Resistor

Cooperative-feedback systems by definition require multiple molecules, i.e., homo-oligomers, to activate. As a result, the transcriptional regulator must first reach a critical concentration before the circuit responds. A feedback resistor might allow a circuit to respond and activate much more quickly than cooperative feedback will (Figure 8A,), which might be an advantage for viral circuits. Interestingly, covalent modification in enzymatic cascades provides increased sensitivity (equivalent to the sensitivity provided by cooperativity) [60] so transactivation circuits may not sacrifice sensitivity by choosing a feedback resistor scheme.

Another potentially important aspect of the feedback resistor is the mechanism's robustness to time delays compared to cooperative feedback. Time delays are well known to produce fold or transcritical (exchange-of-stability) bifurcations in multi-stable systems [61,62]. Cooperative-feedback systems are thus subject to the destabilizing affects of time delays, whereas feedback resistor systems, having only a single steady state, are virtually immune to the destabilizing affects of time delays. This result can also be understood by remembering that the feedback resistor itself has a mathematical formalism with a gamma-distributed time delay [39].

For transactivators with moderate protein or mRNA half-lives, a feedback resistor maintains off state stability across a relatively wide range of kinetic parameters (see stability criterion **ST1**). Animal virus transactivators appear to be moderately long lived, and it is interesting to contrast this aspect of the feedback resistor with the case of long-lived cooperative-feedback molecules. Mathematically, self-cooperative feedback loops (e.g., $dx/dt = k \cdot x^n - \delta x$) are deterministically stable at $x = 0$, but if the circuit is perturbed by noise to values above the unstable equilibrium (the "separatrix" $x = [\delta/k]^{1/n-1}$), then the off state is effectively destabilized; the value of the unstable separatrix $[\delta/k]^{1/n-1}$ decreases as the cooperativity $n$ increases, the off state is thus made increasingly less stable. This argument explains the results of Figure 8A, showing that a feedback resistor off state is more stable than a cooperative-feedback off state. In the feedback resistor case, a fast backward rate ($k_{rev}$) can easily maintain off state stability when δ is small.

There may also be an evolutionary burden for small viral genomes (like HIV-1) employing multi-stable cooperative feedback. All things being equal, in a noisy environment, cooperative feedback would require an additional repressor to detect and restabilize the off state once transactivation is initiated. When cooperative feedback loops are perturbed to values above the unstable equilibrium, then a strong outside control (e.g., negative feedback) is necessary to bring the circuit back into the regime where the off state is again stable and attracting. In our simulations, stochastic fluctuations can easily induce a hypothetical cooperative Tat feedback circuit to transactivate past the unstable equilibrium (Figure 8A). Thus, off state stabilization in the hypothetical cooperative Tat feedback loop would require an additional extrinsic repressor to return the circuit to the off state. The λ-phage Cro transactivator may perform this extrinsic repression function at high Cro concentrations when it binds to $O_{R1}$, $O_{R2}$, and $O_{R3}$, thereby shutting down its own expression [2].

## The Tat Feedback Resistor and HIV Latency

The Tat feedback resistor, its resultant pulse expression, and associated off state stability directly explain our previous observation of stochastic gene expression in HIV-1 Tat driving phenotypic diversity and possibly contributing to HIV-1 latency [34]. After HIV infection of the cell, there would be a stochastic decaying pulse in Tat, which, depending upon its strength and duration, may or may not drive subsequent HIV-1 lifecycle events (e.g., Rev export of unspliced HIV-1 mRNA). Clearly, the action of the feedback resistor in establishing HIV-1 latency may be aided by restriction at the level of transcriptional initiation (via reduced NF-κB activity or chromatin mediated effects) [45] or restriction at the level of translation (via nuclear retention of Tat transcripts) [63].

The HIV-1 Tat feedback resistor model also offers clinical insights, because HIV-1 is known to adopt a latent state in CD4+ memory T cells, and the Tat off state appears necessary and sufficient for maintaining latency of the provirus [27,45]. First, the feedback resistor may offer an explanation as to why two acetylation sites (K28 and K50) are conserved in the Tat protein. Because acetylation of either K28 or K50 can activate Tat, both must be deacetylated to inactivate a Tat molecule. If only a single acetylation site were present in Tat, the transactivator could be more easily inactivated by deacetylation, and this might make the off state too stable. The detection of small numbers of Tat transcripts in latently infected CD4 memory cells from patients with HIV [64] provides some support for this hypothesis and for the feedback resistor mechanism in general. Secondly, HIV patients on highly active anti-retroviral therapy (HAART) exhibit occasional intermittent "blips" of HIV viremia, despite powerful viral repression by HAART. Previous theories have proposed that intermittent viremia is due to immune activation of latent cells (the larger the reservoir, the more blips). But, for a significant group of patients, the frequency of intermittent viral blips is not dependent on the size of the latent reservoir [65,66]. In the stochastic feedback resistor model, intermittent blips of Tat transcription would be predicted to occur in latent cells independent of the latent reservoir size. Our evidence also suggests that viremia blips may result from environmental SirT1 inhibitors, such as the flavoring agent dihydrocoumarin [67], that weaken the Tat feedback resistor.

Finally, an exciting prospect arises upon considering the data presented here in concert with the findings of Pagans et al. [50], who report that a SirT1 inhibitor (HR73) decreases HIV-1 gene expression. Because our data show that HR73 and other inhibitors activate expression of an LTR-GFP-IRES-Tat circuit, destabilize the Tat off state, and increase single-cell transactivation kinetics, it is likely that SirT1 plays different roles at different times in infection. When Tat levels are relatively low, SirT1 may compose part of the feedback resistor. But at later times, in infection SirT1's recycling mechanism may dominate to "turbo-charge" Tat transactivation. Cooperativity between SirT1 and Tat could explain this discrepancy. Together, these data suggest that SirT1 inhibitors may be representative of a new class of anti-HIV drugs that both activate HIV latent cells (by destabilizing the Tat off state) and also reduce subsequent lytic replication from these activated cells.

## Materials and Methods

**Plasmids.** The LTR-GFP, LTR-CMV-GFP, LTR-GFP-IRES-Tat, LTR-Tat-IRES-GFP, LTR-GFP-IRES-Tat(K50A), and LTR-Tat-GFP lentiviral vectors have all been previously described [34], and the EF1α-GFP Ubq-GFP lentivirus vectors were kind gifts from Ronald Weiss (Princeton University, Princeton, New Jersey, United States). SirT1 overexpression retroviral vectors have been previously described [51], and were obtained from Izumi Horikawa (National Institutes of Health, Bethesda, Maryland, United States). Lentiviral and retroviral vectors were packaged in 293T cells by complementation as previously described [68].

**Cell culture, drug perturbations, and sorting.** Cells were maintained at densities between $2 \times 10^5$ and $2 \times 10^6$ cells/ml at 37 °C under $CO_2$ and humidity in RPMI 1640 supplemented with 10% fetal calf serum. The LTR-GFP-IRES-Tat Jurkat clone E7 (a clone sorted from the dim region of GFP fluorescence and having an intergenic integration site) and LTR-GFP Jurkat clones D5 and E11 were used throughout this study. These clones and others we examined (the LGIT clones: D8–2, F5–2, B6–2, E6–1, D10–2, D9–1, C3br, E9br, and E4–1, all having intergenic integration sites) have been previously characterized [34]. Sirtinol and splitomycin were obtained from Calbiochem (Santa Cruz, California, United States), and TNF-α, phorbol myristate-acetate (PMA), nicotinamide, resveratrol, and dihydrocoumarin were obtained from Sigma Chemical (St. Louis, Missouri, United States). All drugs were dissolved in either DMSO or ethanol. DMSO or ethanol controls were included in all experiments, and neither DMSO nor ethanol had significant affects on any of the clones tested. Purified HIV-1 Tat protein was obtained from ABL Inc. (Kensington, Maryland, United States) and cells were exposed as previously described [42]. HR73 was a kind gift from Eric Verdin (University of California, San Francisco, California, United States). Briefly, flow cytometry and FACS parameters were as follows: living cells (in growth media) were gated on forward-versus-side scattering and sorted according to the level of GFP expression. At least 10,000 GFP events were recorded for each experiment. For FACS, at least 100,000 cells were sorted, and post-sort analysis was conducted to verify sorting fidelity. Cytometer and sorter details can be found at http://www.molbio.princeton.edu/facility/flowcyt/cytometers.html.

**Real-time single-cell expression kinetics, homo-FRET, and Western blot analysis.** Jurkat and SupT1 cells were imaged on a Perkin-Elmer UltraView spinning disk confocal microscope fitted with a live-cell chamber (Bioptechs, Butler, Pennsylvania, United States). All experiments were performed at 37 °C under $CO_2$. Cells were immobilized by incubation in glass-bottom cell culture dishes (Matek, Ashland, Massachusetts, United States) for 1 h, drug perturbations were applied, and images were captured every 5–10 min for 12–15 h with an acquisition speed of 100–1,000 ms, depending on the experiment. Cells were tracked and data was analyzed with Perkin-Elmer UltraView software and an in-house MatLab (The Mathworks, Natwick, Massachusetts, United States) code. For each experiment, temporal trajectories of at least 30 single cells were pooled and analyzed (typically, 100–300 cells were captured). Homo-FRET polarization anisotropy microscopy measurements were conducted using a polarizing beam splitter and two-detector system as previously described [69], except that our system was built as a two-photon homo-FRET microscope [49]. Parallel and perpendicular polarization images of immobilized Jurkats were taken at successive z-planes, and r values for each pixel (or region of interest) calculated by the standard $r = (I_P - gI_\perp)/(I_P + 2gI_\perp)$, where $I_P$ and $I_\perp$ refer to parallel and perpendicular intensities, respectively, and the correction factor g varied between 0.98 and 1.02 on our system (the g value needed for setting r=0 upon imaging fluorescein in solution). Images were collected and analyzed using an in-house MatLab code.

For Western blots, we used the FLAG-tag fused to Tat in the LTR-GFP-IRES-Tat vector. LTR-GFP-IRES-Tat cells (clone E7) were exposed to TNF-α at 10 ng/ml and $10^6$ cells were collected at specified time points, washed in phosphate buffered saline (PBS), and immediately frozen at −80 °C. Cell pellets were lysed in 25 μl RIPA buffer, boiled for 5 min in loading buffer containing DTT, run on a 5% SDS-polyacrylamide gel, and transferred to blotting paper via semi-dry method. Blots were blocked, probed using an anti-FLAG M2 antibody (catalog #F-1804, Sigma-Aldrich), and developed by auto-radiography. Blots were then re-blocked, washed, and re-probed with an anti-α-tubulin antibody (catalog #T6199, Sigma-Aldrich) and developed. Blots were quantitatively analyzed on a Molecular

Dynamics Storm system using ImageQuant software (GE Healthcare, Piscataway, New Jersey, United States).

**Mathematical modeling.** ODE modeling and nonlinear least-squares fitting were carried out in Berkeley Madonna software. Direct stochastic simulation of the chemical master equation, using the method of Gillespie, was coded in C and plotted using PLPLOT, as previously described [34], and the reaction scheme was as follows: $Tat + LTR \rightarrow Tat_D$, $Tat_D \rightarrow Tat_A$, $Tat_A \rightarrow Tat_D$, $Tat_A \rightarrow LTR + GFP + Tat$, $Tat \rightarrow 0$, $GFP \rightarrow 0$, where $Tat$ refers to free de-acetylated Tat and $Tat_D$ and $Tat_A$ refer to the LTR bound forms of Tat (rates used where: $k_{basal} = 0.1/s$, $k_{for} = 0.03/s$ or $0.015/s$, $k_{rev} = 40/s$, $k_{TR} = 0.02/s$, $\delta_{Tat} = 0.000016/s$, and $\delta_{GFP} = 0.000004/s$, respectively) with initial conditions: LTR = 1 molecule and all molecular species = 0 molecules. These parameter values satisfy the stability criterion **ST1**. Code available upon request.

## Supporting Information

**Figure S1.** Steady-State and Stability Analysis of the Feedback Resistor Model, Derivation of the Stability Criterion **ST1**, and Difference between the Feedback Resistor and Standard Equilibrium Rates of Transcription

Found at doi:10.1371/journal.pbio.0050009.sg001 (495 KB PDF).

**Figure S2.** Addition of Exogenous Tat Protein to LTR-GFP-IRES-Tat Cells Circumvents the Feedback Circuit and Directly Transactivates the LTR, Activation Kinetics Appear Subexponential/Subquadratic in Time as Evidenced by Plotting on a Log Scale

Found at doi:10.1371/journal.pbio.0050009.sg002 (66 KB PDF).

**Figure S3.** LTR-GFP-IRES-Tat Clone Incubated in Titrating Concentrations of Exogenous Tat Protein

This figure is the analog of Figure 4A.

Found at doi:10.1371/journal.pbio.0050009.sg003 (120 KB PDF).

**Figure S4.** Analytic Solutions to the Differential Equations 1 and 2

Found at doi:10.1371/journal.pbio.0050009.sg004 (106 KB PDF).

**Figure S5.** Comparison of Kinetic Relaxation to Off State for LTR-GFP-IRES-Tat Versus LTR-mRFP-IRES-TatGFP

TatGFP has a ~4-fold shorter half-life than GFP and relaxes ~4 times quicker.

Found at doi:10.1371/journal.pbio.0050009.sg005 (14 KB PDF).

**Video S1.** Single-Cell Confocal Movie of LTR-GFP-IRES-Tat Clone Monitored for 12 h after Addition of TNF-α

The actual run-time is approximately 7 s, conversion to .avi introduced pixel saturation.

Found at doi:10.1371/journal.pbio.0050009.sv001 (934 KB WMV).

## Acknowledgments

We thank David Botstein, Eric Verdin, Jasper Rine, Adam Arkin, David Schaffer, John Burnett, and members of the Shenk lab for helpful discussions and encouragement. Monica Skoge provided the TSR-YFP controls, Stephan Thiberge at the Lewis-Sigler Imaging Core provided invaluable microscopy expertise, Nat Moorman provided technical expertise, and Ron Weiss and Izumi Horikawa kindly provided reagents. We also thank Lynn Enquist, Ned Wingreen, Ted Cox, Ido Golding, Matt Onsum, Ariel Weinberger, Nat Moorman, and Amy Caudy, who critiqued an earlier manuscript draft.

**Author contributions.** LSW conceived and designed the experiments. LSW performed the experiments. LSW analyzed the data. LSW and TS contributed reagents/materials/analysis tools. LSW and TS wrote the paper.

**Funding.** This work was supported by a Lewis Thomas Fellowship (to LSW) and by National Institutes of Health Grant number CA85786 (to TS).

**Competing interests.** The authors have declared that no competing interests exist.

### References

1. Samoilov M, Plyasunov S, Arkin AP (2005) Stochastic amplification and signaling in enzymatic futile cycles through noise-induced bistability with oscillations. Proc Natl Acad Sci U S A 102: 2310–2315.
2. Ptashne M (2004) A genetic switch: Phage lambda revisited. Cold Spring Harbor (New York): Cold Spring Harbor Laboratory Press. 154 pp.
3. Dodd IB, Perkins AJ, Tsemitsidis D, Egan JB (2001) Octamerization of lambda CI repressor is needed for effective repression of P(RM) and efficient switching from lysogeny. Genes Dev 15: 3013–3022.
4. Hasty J, Pradines J, Dolnik M, Collins JJ (2000) Noise-based switches and amplifiers for gene expression. Proc Natl Acad Sci U S A 97: 2075–2080.
5. Brandman O, Ferrell JE Jr., Li R, Meyer T (2005) Interlinked fast and slow positive feedback loops drive reliable cell decisions. Science 310: 496–498.
6. Ragoczy T, Miller G (2001) Autostimulation of the Epstein-Barr virus BRLF1 promoter is mediated through consensus Sp1 and Sp3 binding sites. J Virol 75: 5240–5251.
7. Sarisky RT, Gao Z, Lieberman PM, Fixman ED, Hayward GS, et al. (1996) A replication function associated with the activation domain of the Epstein-Barr virus Zta transactivator. J Virol 70: 8340–8347.
8. Ragoczy T, Heston L, Miller G (1998) The Epstein-Barr virus Rta protein activates lytic cycle genes and can disrupt latency in B lymphocytes. J Virol 72: 7978–7984.
9. Cai W, Astor TL, Liptak LM, Cho C, Coen DM, et al. (1993) The herpes simplex virus type 1 regulatory protein ICP0 enhances virus replication during acute infection and reactivation from latency. J Virol 67: 7501–7512.
10. Roizman B, Gu H, Mandel G (2005) The first 30 minutes in the life of a virus: unREST in the nucleus. Cell Cycle 4: 1019–1021.
11. Kent JR, Kang W, Miller CG, Fraser NW (2003) Herpes simplex virus latency-associated transcript gene function. J Neurovirol 9: 285–290.
12. Marchini A, Liu H, Zhu H (2001) Human cytomegalovirus with IE-2 (UL122) deleted fails to express early lytic genes. J Virol 75: 1870–1878.
13. Heider JA, Bresnahan WA, Shenk TE (2002) Construction of a rationally designed human cytomegalovirus variant encoding a temperature-sensitive immediate-early 2 protein. Proc Natl Acad Sci U S A 99: 3141–3146.
14. Malone CL, Vesole DH, Stinski MF (1990) Transactivation of a human cytomegalovirus early promoter by gene products from the immediate-early gene IE2 and augmentation by IE1: Mutational analysis of the viral proteins. J Virol 64: 1498–1506.
15. Cherrington JM, Mocarski ES (1989) Human cytomegalovirus ie1 transactivates the alpha promoter-enhancer via an 18-base-pair repeat element. J Virol 63: 1435–1440.
16. Cherrington JM, Khoury EL, Mocarski ES (1991) Human cytomegalovirus

17. ie2 negatively regulates alpha gene expression via a short target sequence near the transcription start site. J Virol 65: 887–896.
18. Liu B, Hermiston TW, Stinski MF (1991) A cis-acting element in the major immediate-early (IE) promoter of human cytomegalovirus is required for negative regulation by IE2. J Virol 65: 897–903.
19. Garriga J, Bhattacharya S, Calbo J, Marshall RM, Truongcao M, et al. (2003) CDK9 is constitutively expressed throughout the cell cycle, and its steady-state expression is independent of SKP2. Mol Cell Biol 23: 5165–5173.
20. O'Keeffe B, Fong Y, Chen D, Zhou S, Zhou Q (2000) Requirement for a kinase-specific chaperone pathway in the production of a Cdk9/cyclin T1 heterodimer responsible for P-TEFb-mediated tat stimulation of HIV-1 transcription. J Biol Chem 275: 279–287.
21. Rusche LN, Kirchmaier AL, Rine J (2003) The establishment, inheritance, and function of silenced chromatin in *Saccharomyces cerevisiae*. Annu Rev Biochem 72: 481–516.
22. Thiel G, Lietz M, Hohl M (2004) How mammalian transcriptional repressors work. Eur J Biochem 271: 2855–2862.
23. Kass SU, Landsberger N, Wolffe AP (1997) DNA methylation directs a time-dependent repression of transcription initiation. Curr Biol 7: 157–165.
24. Lassen K, Han Y, Zhou Y, Siliciano J, Siliciano RF (2004) The multifactorial nature of HIV-1 latency. Trends Mol Med 10: 525–531.
25. Mahmoudi T, Parra M, Vries RG, Kauder SE, Verrijzer CP, et al. (2006) The SWI/SNF chromatin-remodeling complex is a cofactor for Tat transactivation of the HIV promoter. J Biol Chem 281: 19960–19968.
26. Ott M, Dorr A, Hetzer-Egger C, Kaehlcke K, Schnolzer M, et al. (2004) Tat acetylation: A regulatory switch between early and late phases in HIV transcription elongation. Novartis Found Symp 259: 182–193.
27. Cullen BR (2003) Nuclear mRNA export: insights from virology. Trends Biochem Sci 28: 419–424.
28. Jordan A, Bisgrove D, Verdin E (2003) HIV reproducibly establishes a latent infection after acute infection of T cells in vitro. Embo J 22: 1868–1877.
29. Laspia MF, Rice AP, Mathews MB (1989) HIV-1 Tat protein increases transcriptional initiation and stabilizes elongation. Cell 59: 283–292.
30. Frankel AD, Bredt DS, Pabo CO (1988) Tat protein from human immunodeficiency virus forms a metal-linked dimer. Science 240: 70–73.
31. Kittiworakarn J, Lecoq A, Moine G, Thai R, Lajeunesse E, et al. (2006) HIV-1 Tat raises an adjuvant-free humoral immune response controlled by its core region and its ability to form cysteine-mediated oligomers. J Biol Chem 281: 3105–3115.
32. Rice AP, Chan F (1991) Tat protein of human immunodeficiency virus type 1 is a monomer when expressed in mammalian cells. Virology 185: 451–454.
33. Stauber RH, Pavlakis GN (1998) Intracellular trafficking and interactions of the HIV-1 Tat protein. Virology 252: 126–136.

33. Tosi G, Meazza R, De Lerma Barbaro A, D'Agostino A, Mazza S, et al. (2000) Highly stable oligomerization forms of HIV-1 Tat detected by monoclonal antibodies and requirement of monomeric forms for the transactivating function on the HIV-1 LTR. Eur J Immunol 30: 1120–1126.

34. Weinberger LS, Burnett JC, Toettcher JE, Arkin AP, Schaffer DV (2005) Stochastic gene expression in a lentiviral positive-feedback loop: HIV-1 Tat fluctuations drive phenotypic diversity. Cell 122: 169–182.

35. Nevels M, Brune W, Shenk T (2004) SUMOylation of the human cytomegalovirus 72-kilodalton IE1 protein facilitates expression of the 86-kilodalton IE2 protein and promotes viral replication. J Virol 78: 7803–7812.

36. Liang Y, Chang J, Lynch SJ, Lukac DM, Ganem D (2002) The lytic switch protein of KSHV activates gene expression via functional interaction with RBP-Jkappa (CSL), the target of the Notch signaling pathway. Genes Dev 16: 1977–1989.

37. Kaehlcke K, Dorr A, Hetzer-Egger C, Kiermer V, Henklein P, et al. (2003) Acetylation of Tat defines a cyclinT1-independent step in HIV transactivation. Mol Cell 12: 167–176.

38. Davido DJ, von Zagorski WF, Lane WS, Schaffer PA (2005) Phosphorylation site mutations affect herpes simplex virus type 1 ICP0 function. J Virol 79: 1232–1243.

39. Mittler JE, Sulzer B, Neumann AU, Perelson AS (1998) Influence of delayed viral production on viral dynamics in HIV-1 infected patients. Math Biosci 152: 143–163.

40. Boutell C, Canning M, Orr A, Everett RD (2005) Reciprocal activities between herpes simplex virus type 1 regulatory protein ICP0, a ubiquitin E3 ligase, and ubiquitin-specific protease USP7. J Virol 79: 12342–12354.

41. Bres V, Kiernan RE, Linares LK, Chable-Bessia C, Plechakova O, et al. (2003) A non-proteolytic role for ubiquitin in Tat-mediated transactivation of the HIV-1 promoter. Nat Cell Biol 5: 754–761.

42. Jordan A, Defechereux P, Verdin E (2001) The site of HIV-1 integration in the human genome determines basal transcriptional activity and response to Tat transactivation. Embo J 20: 1726–1738.

43. Richter S, Ping YH, Rana TM (2002) TAR RNA loop: A scaffold for the assembly of a regulatory switch in HIV replication. Proc Natl Acad Sci U S A 99: 7928–7933.

44. Karn J (2000) Tat, a novel regulator of HIV transcription and latency. In: Kuiken C, McCutchan F, Foley B, editors. HIV sequence compendium 2000. Los Alamos, (New Mexico): Theoretical biology and biophysics group, Los Alamos National Laboratory. pp. 2–18.

45. Williams SA, Chen LF, Kwon H, Ruiz-Jarabo CM, Verdin E, et al. (2006) NF-kappaB p50 promotes HIV latency through HDAC recruitment and repression of transcriptional initiation. Embo J 25: 139–149.

46. Strogatz SH (1994) Nonlinear dynamics and chaos. Cambridge (Massachusetts): Westview Press. 498 p.

47. Kiernan RE, Vanhulle C, Schiltz L, Adam E, Xiao H, et al. (1999) HIV-1 tat transcriptional activity is regulated by acetylation. Embo J 18: 6106–6118.

48. Tramier M, Piolot T, Gautier I, Mignotte V, Coppey J, et al. (2003) Homo-FRET versus hetero-FRET to probe homodimers in living cells. Methods Enzymol 360: 580–597.

49. Volkmer A, Subramaniam V, Birch DJ, Jovin TM (2000) One- and two-photon excited fluorescence lifetimes and anisotropy decays of green fluorescent proteins. Biophys J 78: 1589–1598.

50. Pagans S, Pedal A, North BJ, Kaehlcke K, Marshall BL, et al. (2005) SIRT1 regulates HIV transcription via Tat deacetylation. PLoS Biol 3: e41. doi:10.1371/journal.pbio.0030041.

51. Michishita E, Park JY, Burneskis JM, Barrett JC, Horikawa I (2005) Evolutionarily conserved and nonconserved cellular localizations and functions of human SIRT proteins. Mol Biol Cell 16: 4623–4635.

52. Howitz KT, Bitterman KJ, Cohen HY, Lamming DW, Lavu S, et al. (2003) Small molecule activators of sirtuins extend Saccharomyces cerevisiae lifespan. Nature 425: 191–196.

53. Deng L, de la Fuente C, Fu P, Wang L, Donnelly R, et al. (2000) Acetylation of HIV-1 Tat by CBP/P300 increases transcription of integrated HIV-1 genome and enhances binding to core histones. Virology 277: 278–295.

54. Fall CP (2002) Computational cell biology. New York: Springer. 468 p.

55. Gillespie DT (1977) Exact stochastic simulation of coupled Cemical-reactions. J Phys Chem 81: 2340–2361.

56. Perelson AS, Neumann AU, Markowitz M, Leonard JM, Ho DD (1996) HIV-1 dynamics in vivo: virion clearance rate, infected cell life-span, and viral generation time. Science 271: 1582–1586.

57. Liao W, Tang Y, Kuo YL, Liu BY, Xu CJ, et al. (2003) Kaposi's sarcoma-associated herpesvirus/human herpesvirus 8 transcriptional activator Rta is an oligomeric DNA-binding protein that interacts with tandem arrays of phased A/T-trinucleotide motifs. J Virol 77: 9399–9411.

58. Carrozza MJ, Li B, Florens L, Suganuma T, Swanson SK, et al. (2005) Histone H3 methylation by Set2 directs deacetylation of coding regions by Rpd3S to suppress spurious intragenic transcription. Cell 123: 581–592.

59. Keogh MC, Kurdistani SK, Morris SA, Ahn SH, Podolny V, et al. (2005) Cotranscriptional set2 methylation of histone H3 lysine 36 recruits a repressive Rpd3 complex. Cell 123: 593–605.

60. Goldbeter A, Koshland DE Jr. (1981) An amplified sensitivity arising from covalent modification in biological systems. Proc Natl Acad Sci U S A 78: 6840–6844.

61. Murray JD (2002) Mathematical biology II. New York: Springer. 736 pp.

62. Kuang Y (1993) Delay differential equations: With applications in population dynamics. Boston: Academic Press. 398 pp.

63. Lassen KG, Ramyar KX, Bailey JR, Zhou Y, Siliciano RF (2006) Nuclear retention of multiply spliced HIV-1 RNA in resting CD4+ T cells. PLoS Pathog 2: e68. doi:10.1371/journal.ppat.0020068.

64. Lassen KG, Bailey JR, Siliciano RF (2004) Analysis of human immunodeficiency virus type 1 transcriptional elongation in resting CD4+ T cells in vivo. J Virol 78: 9105–9114.

65. Ramratnam B, Ribeiro R, He T, Chung C, Simon V, et al. (2004) Intensification of antiretroviral therapy accelerates the decay of the HIV-1 latent reservoir and decreases, but does not eliminate, ongoing virus replication. J Acquir Immune Defic Syndr 35: 33–37.

66. Ramratnam B, Mittler JE, Zhang L, Boden D, Hurley A, et al. (2000) The decay of the latent reservoir of replication-competent HIV-1 is inversely correlated with the extent of residual viral replication during prolonged anti-retroviral therapy. Nat Med 6: 82–85.

67. Olaharski AJ, Rine J, Marshall BL, Babiarz J, Zhang L, et al. (2005) The flavoring agent dihydrocoumarin reverses epigenetic silencing and inhibits sirtuin deacetylases. PLoS Genet 1: e77. doi:10.1371/journal.pgen.0010077.

68. Dull T, Zufferey R, Kelly M, Mandel RJ, Nguyen M, et al. (1998) A third-generation lentivirus vector with a conditional packaging system. J Virol 72: 8463–8471.

69. Vaknin A, Berg HC (2006) Osmotic stress mechanically perturbs chemoreceptors in Escherichia coli. Proc Natl Acad Sci U S A 103: 592–596.

70. Hoffmann A, Levchenko A, Scott ML, Baltimore D (2002) The IkappaB-NF-kappaB signaling module: Temporal control and selective gene activation. Science 298: 1241–1245.